\documentclass[sigconf]{acmart}

\AtBeginDocument{%
  \providecommand\BibTeX{{%
    \normalfont B\kern-0.5em{\scshape i\kern-0.25em b}\kern-0.8em\TeX}}}

\usepackage{multicol}
\usepackage{graphicx}
\usepackage{color}
\usepackage{float}

\restylefloat{figure}
\usepackage{graphicx}
\usepackage{caption}
\usepackage{subcaption}

\usepackage{enumitem}
\usepackage{multirow}

\copyrightyear{2019} 
\acmYear{2019} 
\setcopyright{acmlicensed}
\acmConference[IDEAS'19]{23rd International Database Engineering & Applications Symposium}{June 10--12, 2019}{Athens, Greece}
\acmBooktitle{23rd International Database Engineering \& Applications Symposium (IDEAS'19), June 10--12, 2019, Athens, Greece}
\acmPrice{15.00}
\acmDOI{10.1145/3331076.3331095}
\acmISBN{978-1-4503-6249-8/19/06}

\begin{document}

\title{Classification of eye-state using EEG recordings: speed-up gains using signal epochs and mutual information measure}

\author{Phoebe M Asquith}
\affiliation{%
  \institution{Cardiff University}
  \streetaddress{Park Place}
  \city{Cardiff}
  \postcode{CF10 3AT}
  \country{UK}}
\email{asquithpm@cardiff.ac.uk}

\author{Hisham Ihshaish}
\affiliation{%
  \institution{University of the West of England}
  \streetaddress{Computer Science Research Centre (CSRC)}
  \city{Bristol}
  \country{UK}}
\email{hisham.ihshaish@uwe.ac.uk}

\renewcommand{\shortauthors}{Asquith and Ihshaish}%%
\renewcommand{\shorttitle}{Speed-up gains using signal epochs and mutual information measure}

%% article.] 
\begin{abstract}
The classification of electroencephalography (EEG) signals is useful in a wide range of applications such as seizure detection/prediction, motor imagery classification, emotion classification and drug effects diagnosis, amongst others. With the large number of EEG channels acquired, it has become vital that efficient data-reduction methods are developed, with varying importance from one application to another. It is also important that online classification is achieved during EEG recording for many applications, to monitor changes as they happen. In this paper we introduce a method based on Mutual Information (MI), for channel selection. Obtained results show that whilst there is a penalty on classification accuracy scores, promising speed-up gains can be achieved using MI techniques. Using MI with signal epochs (3secs) containing signal transitions enhances these speed-up gains. This work is exploratory and we suggest further research to be carried out for validation and development. Benefits to improving classification speed include improving application in clinical or educational settings.  
\end{abstract}

\begin{CCSXML}
<ccs2012>

<concept>

<concept_id>10002950.10003624.10003633</concept_id>
 <concept_desc>Mathematics of computing~Graph theory</concept_desc>

<concept_significance>500</concept_significance>
</concept>

<concept>

<concept_id>10002950.10003648.10003688.10003693</concept_id>
 <concept_desc>Mathematics of computing~Time series analysis</concept_desc>

<concept_significance>500</concept_significance>
</concept>

<concept>

<concept_id>10010405.10010455.10010459</concept_id>
 <concept_desc>Applied computing~Psychology</concept_desc>

<concept_significance>500</concept_significance>
</concept>

<concept>

<concept_id>10010583.10010588.10010596</concept_id>
 <concept_desc>Hardware~Sensor devices and platforms</concept_desc>

<concept_significance>500</concept_significance>
</concept>
</ccs2012>

\end{CCSXML}

 \ccsdesc[500]{Mathematics of computing~Graph theory}
 \ccsdesc[500]{Mathematics of computing~Time series analysis}
 \ccsdesc[500]{Applied computing~Psychology}
 \ccsdesc[500]{Hardware~Sensor devices and platforms}

\copyrightyear{2019} 
\acmYear{2019}

\keywords{electroencephalogram (EEG) analysis, eye-blink detection, time series analysis, graph theory applications, psychology. mutual information measure}


\maketitle

\section{Introduction}
\label{sect:introduction}

Since its invention in 1929 \cite{Berger1931}, the electroencephalogram (EEG) has allowed the recording and interpretation of the electro-magnetic activity of neurons, from the scalp. Research using this technology has allowed crucial insights into the sleep wake cycle (e.g. \cite{Ferrara2011}), neuropsychological abnormality (e.g. \cite{Stam2005}), functional networks in the brain (e.g. \cite{BROVELLI2002130}) and neural development (e.g. \cite{Boersma:2011aa}). 

Recently, identifying eye-state using EEG has become of interest with findings that eye-state behavior such as blink frequency can demonstrate stress response \cite{Haak2009} or an underlying neuropsychological problem \cite{Sandyk:1990aa}. EEG signal changes related to eye-state have often been identified by separating raw data into different frequency bands \cite{Vermani:2015aa}. However, this does not allow for online classification of eye-state.

\begin{figure}[H]
\begin{center}

\includegraphics[width=\linewidth]{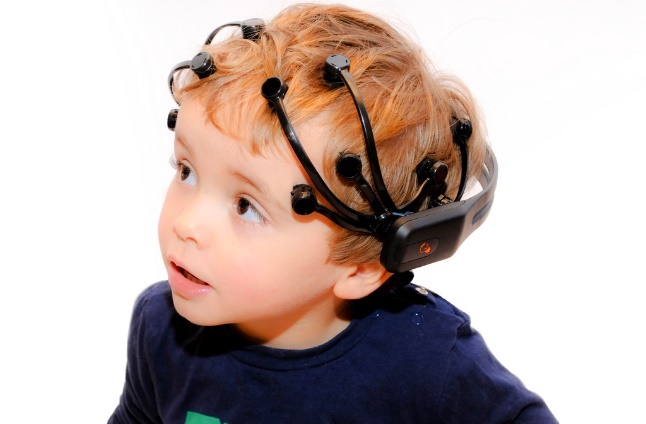}
\caption{Example of portable EEG use with children.}
\label{fig:child}
\end{center}
\end{figure}

More recently the use of portable EEGs has become more prevalent, with the development of innovative technologies (see Fig. \ref{fig:child}). Research has demonstrated that with use of portable headsets, the eye-state of a participant can be identified using the raw time-series recorded at different channels, rather than separating data into different frequency bands \cite{Wang:2014aa}. Despite some concerns around the measurement capabilities of the headsets, the potential of portable devices in current and future research is recognized within the field (e.g. see \cite{XU2018340} for review in educational research). Portable EEGs are easier to implement than traditional EEGs and can be used with subjects "in the field" or who may have difficulty sitting still (e.g. young children). Having online eye-state classification capabilities with this portable technology is an exciting step towards a dynamic resource in cognitive-neuroscientific research. 

The application of machine learning methods for the classification of EEG signals has been widely explored in the last two decades. Examples include methods for feature selection and optimization as in \cite{BAIG2017184} and channel selection as in \cite{YANG2012117, Alotaiby2015}, amongst others (see \cite{Rsler2013AFS} for a wide range of machine learning methods applied). 

In previous work EEG signals have been used to classify eye-state relatively successfully using Incremental Attribute Learning (IAL) with extended timeseries \cite{Wang:2014aa}. Epochs of ten seconds have also been adequate for identifying drowsiness from eyestate \cite{YEO2009115}. However, to be useful as an online classifier, a shorter snapshot of data must be used to identify eye-state rather than an extended time-series, to reduce calculation time and processing power. This is also important for identifying blinks, which typically last 100-400ms \cite{Reddy:2016aa}. Levy \cite{Levy:1987aa} explored the effect of epoch length on signal analysis of the EEG and found that epochs as short as 2 seconds could be used for intraoperative EEG monitoring. For eye-state classification in particular, it has been demonstrated that a snapshot of EEG signal time-series in the alpha frequency range can be used to identify eye-state, rather than an extended time-series \cite{Barry:2007aa}.

In this research, we provide experimental analysis for sample size reduction based on a method to capture signals in discrete EEG signal slices compared to longer EEG signal time-series. Additionally, we investigate the effect of possible signal redundancy on classification scores and computational performance. This will be investigated using the raw EEG data rather than splitting it into different frequency bands, therefore eliminating data preparation steps.

Results show that with both channel selection and sample reduction methods, we could accomplish comparable classification results with KNN, Support Vector Machines (Classifier: SVC), KNN and RF when run on the entire dataset containing signals from all channels. Additionally, outcomes suggest that significant computational speed-up could be achieved using a Mutual Information (MI) measure for EEG 

\section{Dataset}
\label{sec:Data}

\begin{figure}[!]
\begin{center}

\includegraphics[width=.8\columnwidth]{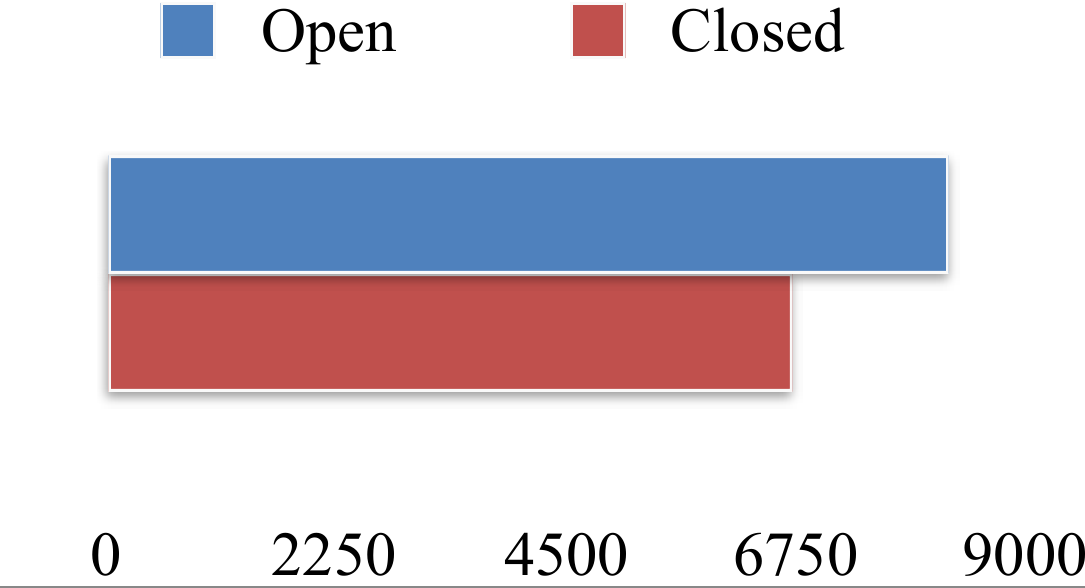}
\caption{Eye-state distribution of the recorded observations.}
\label{fig:dist}
\end{center}
\end{figure}

The data corpus explored in this work was collected and compiled by Roesler \cite{Rsler2013AFS}, and provided for open access on UCI data repository\cite{UCI}. The dataset comprises of raw electro-magnetic recordings taken from the scalp of one participant and information about eye-state (eyes open or closed) over the same time period. The participant was asked to relax, look forwards towards a camera and blink naturally, without restriction \cite{Garrett:2003aa}. While looking toward the camera a video was recorded of the eye. Once recorded the video data was coded; binary labels were used to identify the two different eye-states; '1' for an "eye-blink" and '0' for "eye-open" state — the distribution of eye-states over the course of the recording in the dataset is shown in Fig. \ref{fig:dist}.

\begin{figure}[!]
\begin{center}

\includegraphics[width=.9\columnwidth]{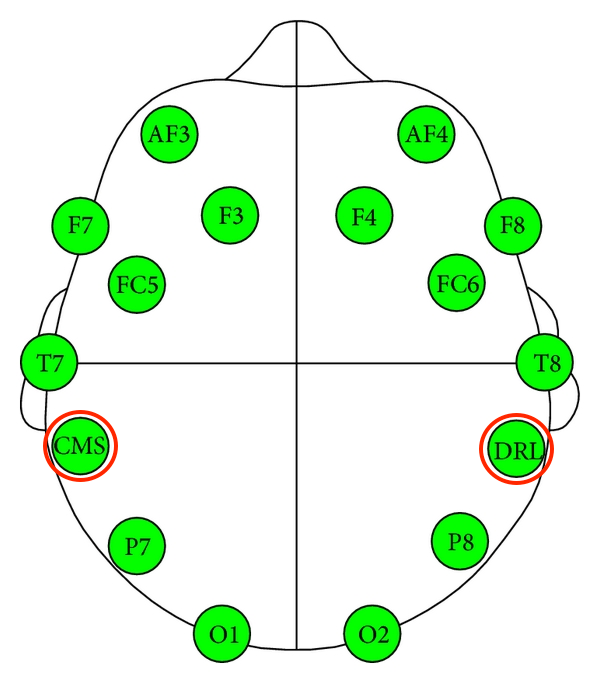}
\caption{The 14 EEG channels compiled by Roesler\cite{Rsler2013AFS}. Excluded channels are circled in red. }
\label{fig:channles}%
\end{center}
\end{figure}

During this time, recordings were also taken at the scalp using the Emotiv EEG Neuroheadset, which measured the electro-magnetic signal at 14 electrode positions (see Fig. \ref{fig:channles} - note that two electrode positions where excluded, as indicated in the figure). 14980 sequential timepoints (observations) were recorded from each of the 14 EEG channels (features), The recording took place over 117 seconds period (this is a rate of 128Hz) and measured signals were stored as floating-point values.

\begin{figure*}%
\centering
\begin{subfigure}{.5\linewidth}
\includegraphics[width=.9\linewidth]{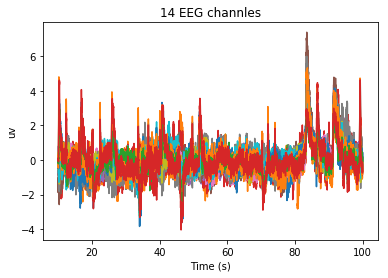}%
\caption{14 channel time-series}
   \label{fig:ts}
\end{subfigure}\hfill%
\begin{subfigure}{.5\linewidth}
 \includegraphics[width=.9\linewidth]{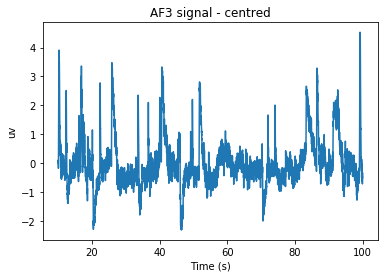}%
 \caption{1 channel time-series}
 \label{fig:af3}
\end{subfigure}\hfill%
\caption{(a) Sliced window of the 14 EEG channel centered time-series- um is signal voltage. Signals show similarity across the EEG channels. (b) Only one channel (AF3) signal is shown.}
\label{fig:testo}
\end{figure*}

Initial exploration of the dataset indicated three outliers (value > 10x the average recording) , which were removed. Observations were therefore reduced to 14977 at each electrode; each of these 14 timeseries represented the signal variability of an electrode over the experimental period. The timeseries were then normalised using zero centring ("de-meaning" applied) to explore the positive and negative deviation from their mean, as an indicator of signal similarity — centred signals are shown in Fig.\ref{fig:testo} . By eye, the timeseries show overall similarity across the different electrodes.

\begin{figure*}%
\centering
\begin{subfigure}{.5\linewidth}
  \includegraphics[width=.9\linewidth]{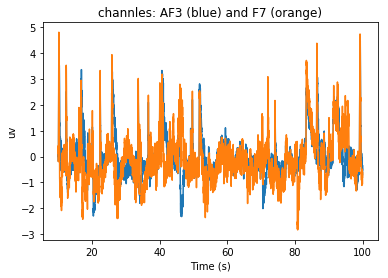}
   \caption{}
   \label{fig:two100}
\end{subfigure}\hfill%
\begin{subfigure}{.5\linewidth}
 \includegraphics[width=.9\linewidth]{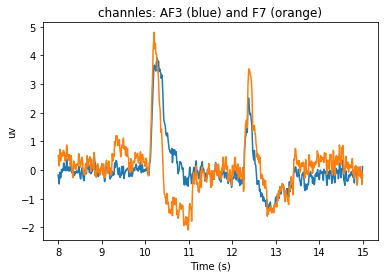}
  \caption{}
   \label{fig:two}
\end{subfigure}\hfill%
\caption{Signal time-series for AF3 and F7, (b) shows a smaller time-slot of the same time-series.}
\label{fig:two-sig}
\end{figure*}

For example, a strong signal similarity can be observed when looking at the time-series of AF3 and F7 as in Fig. \ref{fig:two-sig}.

\begin{figure*}%
\centering
\begin{subfigure}{.5\linewidth}
 \includegraphics[width=.9\linewidth]{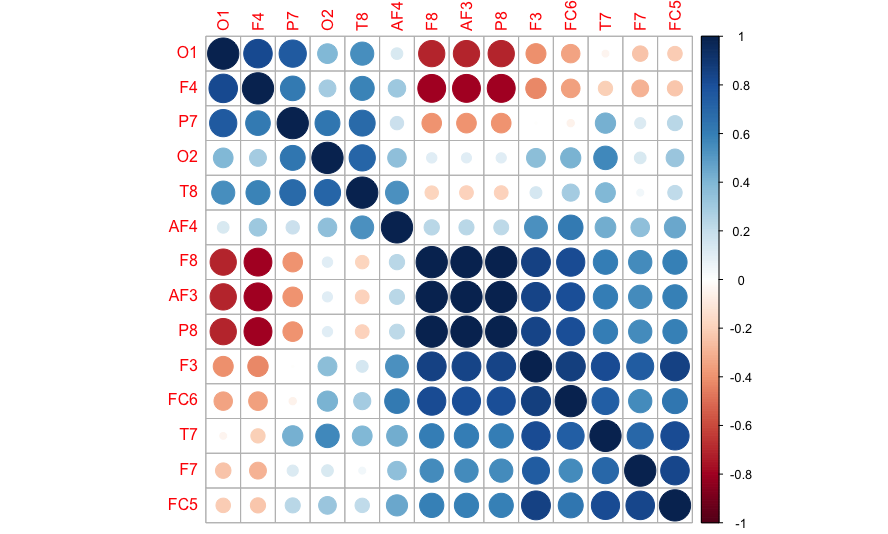}
   \caption{eye-state=1}
   \label{fig:Pos}
\end{subfigure}\hfill%
\begin{subfigure}{.5\linewidth}
\includegraphics[width=.9\linewidth]{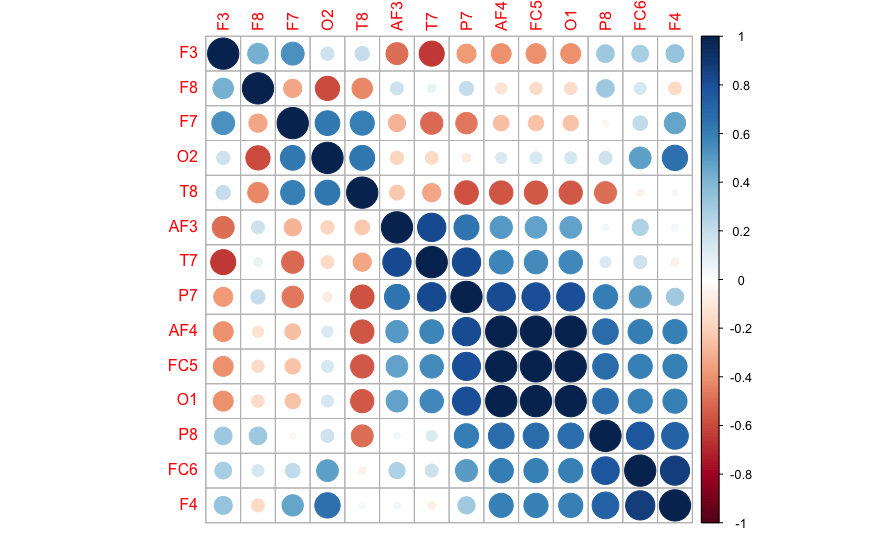}
  \caption{eye-state=0}
  \label{fig:Neg}
\end{subfigure}\hfill%

\caption{Correlation matrices for an 'eye-blink' time-series in (a), and 'eye-open' in (b). Hierarchical clustering \cite{hclut} is applied to cluster higher-correlated channels together. Note the difference of order in each matrix.}
\label{fig:corr}
\end{figure*}

Similarities across the EEG timeseries are observable overall. Indeed, if we separate all EEG time-series relative to eye-state and cross-correlate, topological patterns in signal variability across the channels exist, see Fig. \ref{fig:corr} (note that order of variables varies across the matrices to facilitate visualisation of possible similarities). The correlation between electrode signals is also seen to change during eyes-close state compared to eyes-open.

\section{Methods and findings} 

\begin{figure} [H]
  \begin{subfigure}{0.45\linewidth}
  \includegraphics[width=\linewidth]{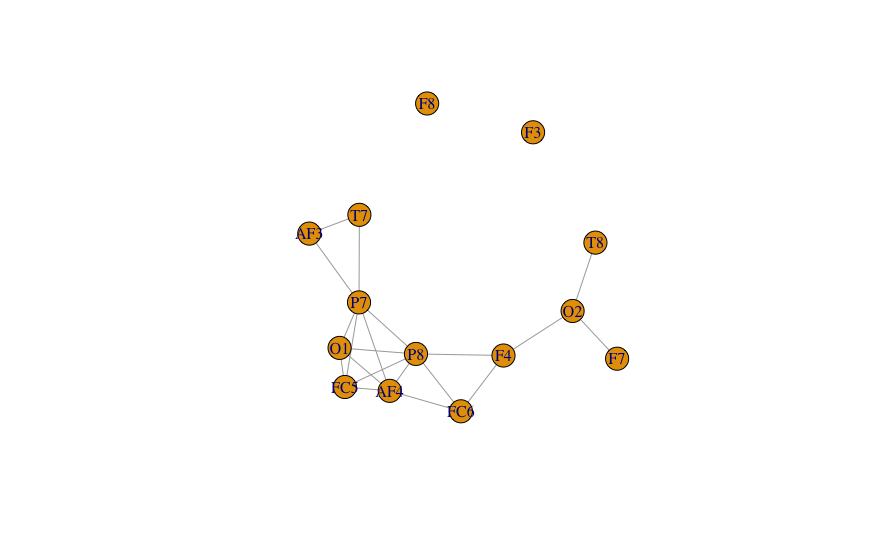}
   \caption{eye-open, $\Theta$ = 0.6 }
  \end{subfigure}
   \begin{subfigure}{0.45\linewidth}
  \includegraphics[width=\linewidth]{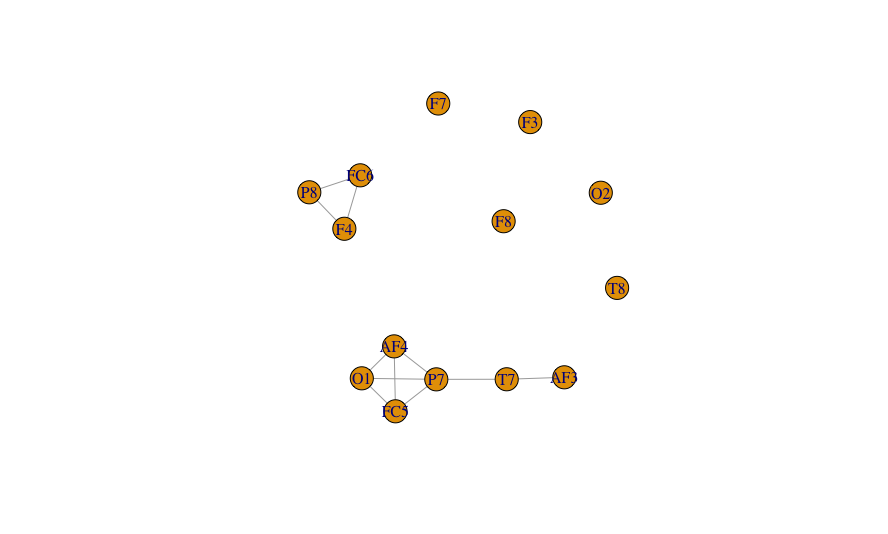}
   \caption{eye-open, $\Theta$ = 0.7 }
  \end{subfigure}
    \hfill
  \begin{subfigure}{0.45\linewidth}
  \includegraphics[width=\linewidth]{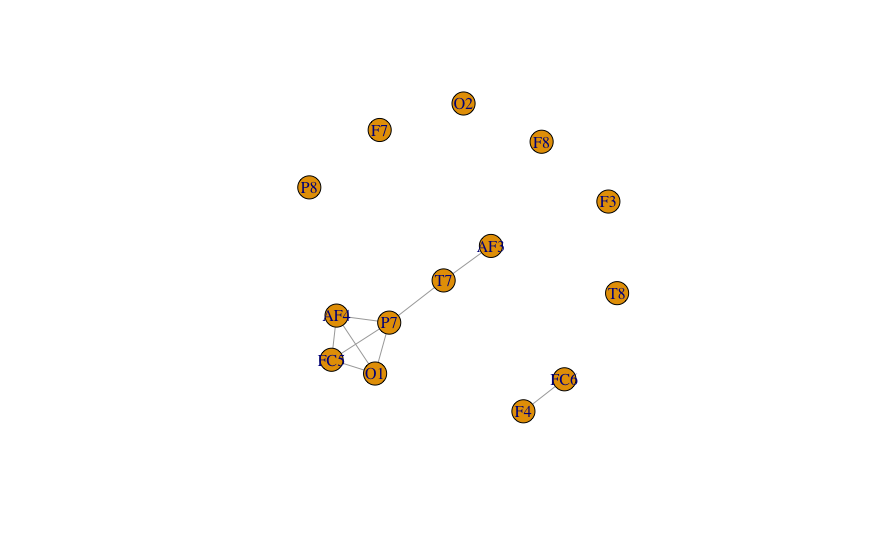}
   \caption{eye-open, $\Theta$ = 0.8}
  \end{subfigure}
  
    \begin{subfigure}{0.45\linewidth}
  \includegraphics[width=\linewidth]{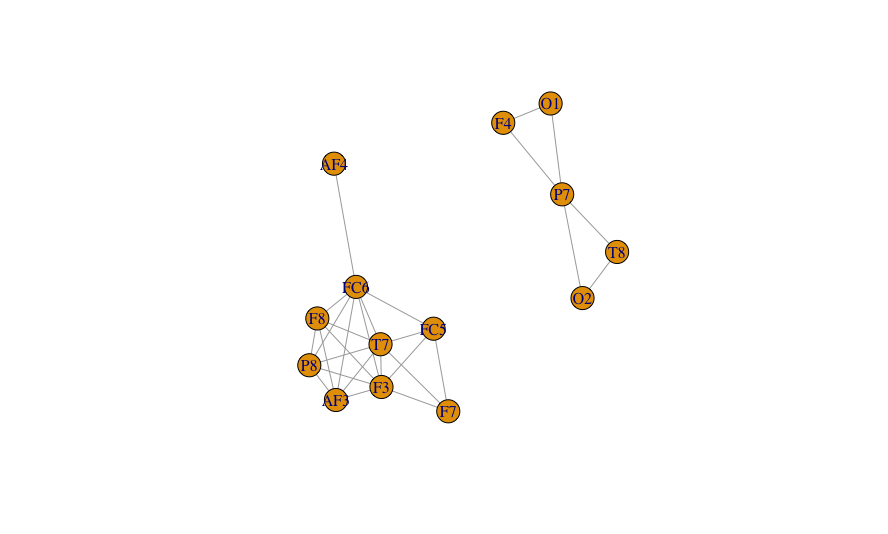}
   \caption{eye-blink, $\Theta$ = 0.6 }
  \end{subfigure}
  
    \hfill
    
   \begin{subfigure}{0.45\linewidth}
  \includegraphics[width=\linewidth]{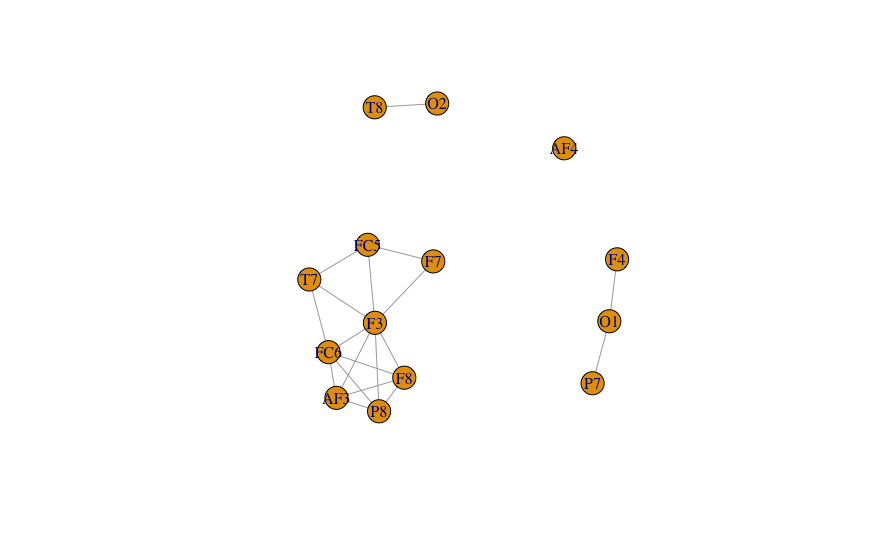}
   \caption{eye-blink, $\Theta$ = 0.7 }
  \end{subfigure}
  \begin{subfigure}{0.45\linewidth}
  \includegraphics[width=\linewidth]{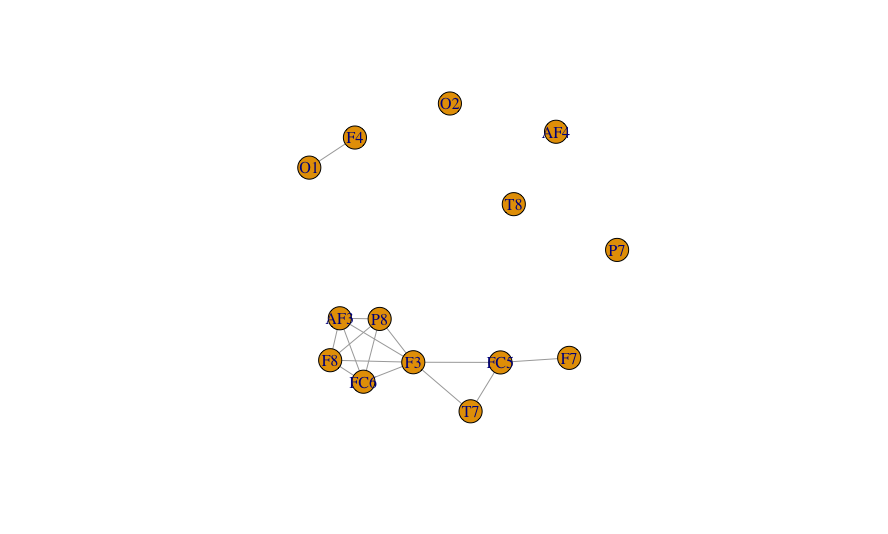}
   \caption{eye-blink, $\Theta$ = 0.8 }
  \end{subfigure}  
  \caption{Different graphs constructed from 14 EEG channel time-series, relative to eye-state and a value for $\Theta$; strength of linear similarity.}
  \label{fig:graphs}
  \end{figure}

Patterns between EEG channel signals relative to eye-state could be further explored by techniques from graph theory. Linear correlations between the time series $T_i(t_k)$ and $T_j(t_k)$ (the Pearson correlation coefficient $R_{ij}$)  given by  
\begin{equation} \label{eq:pearson}
R_{ij} = \frac{ {\sum\limits_{k=1} ^{L} } T_i(t_k) T_j(t_k) } 
{\sqrt{(\sum\limits_{k=1} ^{L}T^2_i(t_k)) (\sum\limits_{k=1} ^{L} T^2_j(t_k))}}
\end{equation}
is widely used \cite{Wilks2011}, whereby strong linearity between two channels can be expressed as a link between two graph nodes. Having derived the correlation matrix $C$, a threshold $\tau$ is usually applied to define strong similarities between graph nodes as `links'. The adjacency matrix $A$ for the graph is then found by
\begin{equation} \label{eq:Adj}
A_{ij} = A_{ji}= \Theta{(C_{ij}-\tau)} - \delta_{ij},
\end{equation}
where $\Theta$ is the Heaviside function and $\delta$ is Kronecker delta. Graphs based on the two different eye-states have been constructed, considering different values for  $\Theta$ — see Fig. \ref{fig:graphs}. 

The constructed graphs (small, provided the number of channels) show a strong dissimilarity in topological structure between first set (eye-open signals) and the second (constructed from eye-blink signals). Studying metrics such as the average degree for nodes of both types of graphs can be used to quantify the topological similarity further. 

\begin{figure}[H]
\begin{center}
\includegraphics[width=\linewidth]{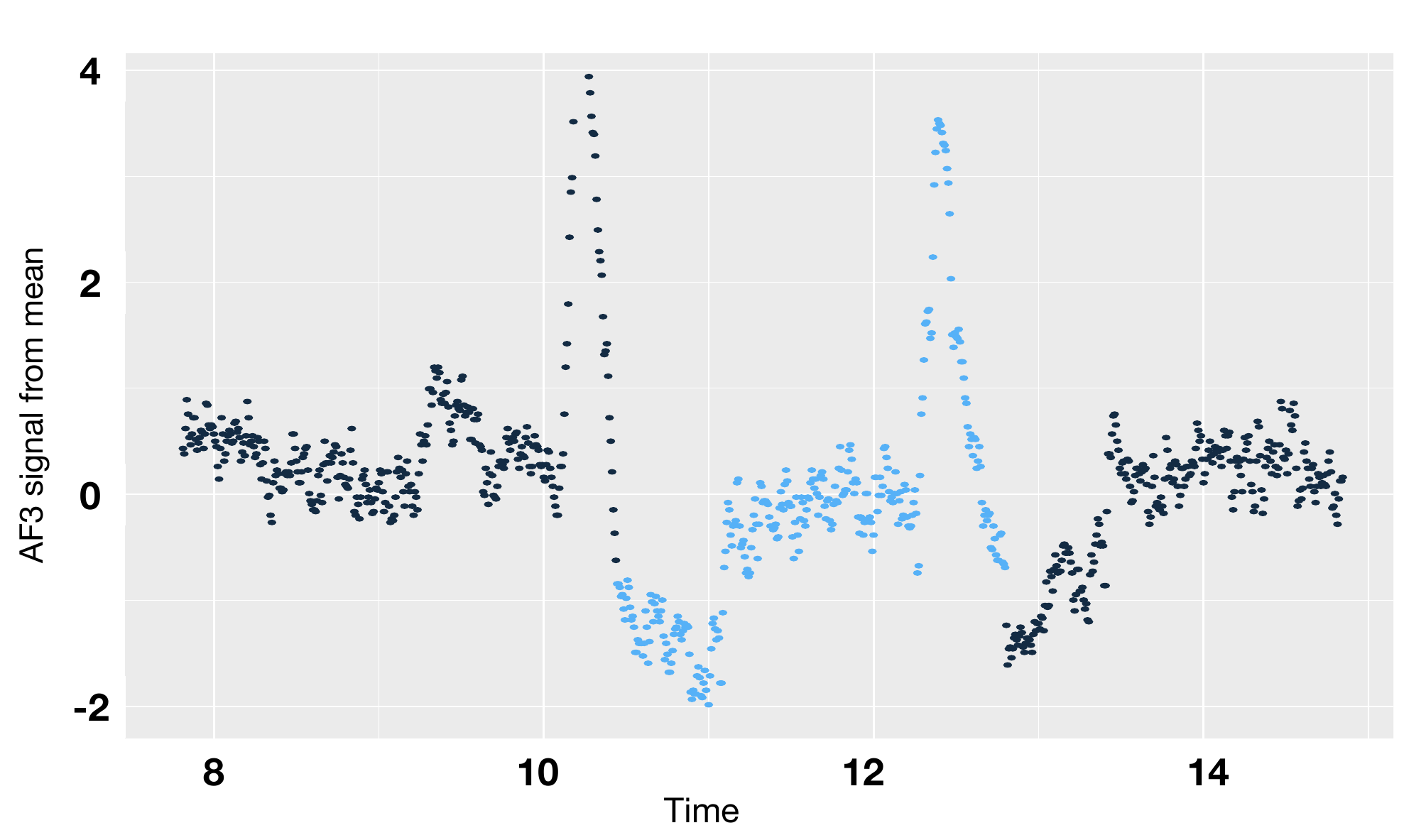}
\caption{Example of F7 signal time-series window — window here is of 7 seconds. Light blue represent recordings in the closed-eye state and dark blue eyes-open.}
\label{fig:slices}
\end{center}
\end{figure}

Based on observed similarities (linear similarity explored here), we argue that machine learning methods should provide comparable results if, on the one hand features' space is reduced based on the relevance of features and their 'score' of redundancy, and on another, the similarity between signals can be captured in shorter time-series of signals. To test both assumptions we sliced the provided time-series for all channels into time-windows of 3 seconds (384 timepoints, collected at a rate of 128Hz), containing a transition between eye-blink and eye-open and vice versa. — Fig \ref{fig:slices} shows a time-series window of 7 seconds for demonstration purposes.  

20 time-series slices (windows of 3s length each) were generated for each channel, resulting in the total number of observations reduced to 7,680. We then implemented a filtering approach based on mutual information ($M_{ij}$), given by

\begin{equation} \label{eq:MI}
M_{ij} = \sum\limits_{T_i, T_j} {P_{ij}(T_i, T_j) \; \log \frac{P_{ij}(T_i, T_j)}{P_{i}(T_i)P_{j}(T_j)}}. 
\end{equation}
Here  $P_{i}(T_i)$ is the probability density function (PDF) of time series $T_i$, 
and  $P_{ij}(T_i, T_j)$ is the joint PDF for $(T_i, T_j)$. 

\begin{figure}[H]
\begin{center}
\includegraphics[width=\linewidth]{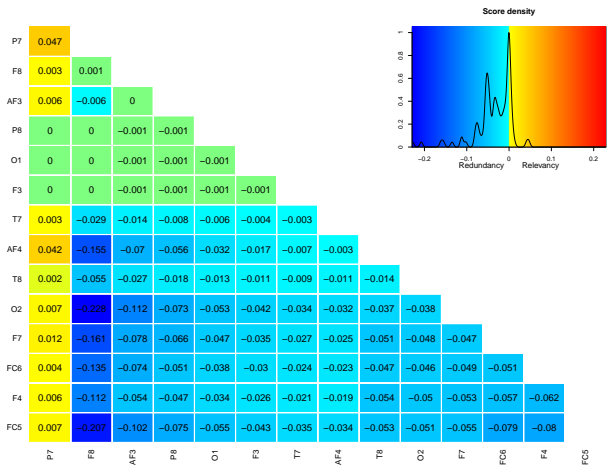}
\caption{ The variable selected at each step is the one with the highest scores (the variables are ordered in the plot). The scores at selection can thus be read from the diagonal. }
\label{fig:varnak}
\end{center}
\end{figure}

The minimum redundancy maximum relevance (mRMRe) algorithm \cite {Battiti:1994aa}, as a filtering method, uses differences of $M_{ij}$ to compute the degree of dependency between multiple random variables. The method then sequentially compares the relevancy/redundancy balance of information between variables, providing scores for both their relevance and redundancy. The variable (channel time-series) selected at each step is the one with the highest score. A negative score indicates a redundancy final trade of information and a positive score indicates a relevancy final trade of information. The scoring results averaged for the 20 time-series slices is shown in Fig \ref{fig:varnak}.

Accordingly we ran a series of experiments on a 3.2 GHz Intel Core i5 processor with 16 GB 1600 MHz DDR3 memory, which included a base run of different machine learning methods for classification; Nearest neighbors (KNN), Logistic Regression, Support Vector Machines classifier (SVC) and Random Forests (RF). The base run included the tuning of K for nearest neighbors and regularisation, using grid search methods, for SVC hyper-parameters (C=10, gamma=0.001) with radial basis function (rbf) kernel. RF is run with bootstrapping enabled and the number of selected features set to 'automatic', in order to decrease variance amongst constructed member trees. 

K-fold validation score of F1 metric was obtained over the entire dataset of 14977 time-series (117 seconds and sampling rate of 128Hz). Although F1 score's value is not the main concern here, a comparison between classification methods' accuracy can be found in \cite{Rsler2013AFS} and in \cite{Reddy:2016aa} for deep learning architectures' performance. The resulting classification F1 Score and processing speed where then compared to those of 3 different experimental settings, for KNN, Logistic Regression, SVC and RF, which included:

\begin{itemize}
\item (A) 9 features from mRMRe analysis are selected from the available 14. Entire dataset of 14977 time-series is used to learn the classifiers. Accuracy and performance results compared to base run are shown in Table 1 below:
	\begin{table}[H]
	\begin{center}
    \begin{tabular}{| l | l | l | l | l |}
    \hline
    Classifier & F1 Score gain & Speed-up gain \\ \hline
    \hline
    $KNN$ &  -0.2 & $2.1x$ \\ \hline
    $LogReg$ & -0.36 & $2x$ \\ \hline
    $SVC$ & -0.3 & $3x$  \\ \hline
    $RF$ & -0.5 & $0.3x$  \\ \hline
    \end{tabular}
    \label{tab:results}
    \caption{ Accuracy and processing time comparison between base run and experiment A. }
\end{center}
\end{table}

\item (B) 9 features from mRMRe analysis are selected from the available 14. 7,680 observations (based reducing time-series to 20 sliced windows containing eye-state transition) used for learning. Accuracy and performance results compared to base run are shown in Table 2 below:
\begin{table}[H]
	\begin{center}
    \begin{tabular}{| l | l | l | l | l |}
    \hline
    Classifier & F1 Score gain & Speed-up gain \\ \hline
    \hline
    $KNN$ &  -0.4 & $4.3x$ \\ \hline
    $LogReg$ & -0.6 & $2.5x$ \\ \hline
    $SVC$ & -0.7 & $5.6x$  \\ \hline
    $RF$ & -0.7 & $3x$  \\ \hline
    \end{tabular}
    \label{tab:resultsb}
    \caption{ Accuracy and processing time comparison between base run and experiment B. }
\end{center}
\end{table}
\item (C) 14 features (no channel selection) on the 7,680 observations are used to learn a classifier. Accuracy and performance results compared to base run are shown in Table 3 below:
\begin{table}[H]
	\begin{center}
    \begin{tabular}{| l | l | l | l | l |}
    \hline
    Classifier & F1 Score gain & Speed-up gain \\ \hline
    \hline
    $KNN$ &  -0.1 & not observed \\ \hline
    $LogReg$ & -0.17 & $2.3x$ \\ \hline
    $SVC$ & -0.3 & $1.9x$  \\ \hline
    $RF$ & -0.63 & $3x$  \\ \hline
    \end{tabular}
    \label{tab:resultsC}
    \caption{ Accuracy and processing time comparison between base run and experiment C.}
\end{center}
\end{table}
\end{itemize}

The obtained results show some classification score penalty in most runs of the presented methods, generally, yet speed-up gain is promising. The accuracy score declines more as both feature reduction as well as data slicing method are applied together, however, processing speed-up gains are maximised. Further tuning to, namely, SVC is believed to show better scores by the application of this work's methods, which we aim at focusing on for future developments. That said, the method introduced can be vastly useful for the analysis of higher-dimensional EEG/MEG signals which are typically characterised by the existence of both redundancy of information in their signals, and most importantly, noise. 

\section{Conclusions and further work}
This paper presented a brief literature on the analysis of the electroencephalogram (EEG) signals and the application of their analysis. Information obtained by Emotiv headsets on subjects/humans include signal time-series from different electrodes, which typically exhibit variability, in response to events designed for a study of interest. The resolution of collected data as well as the quantity of time-series which could be obtained by such devices is, increasingly, producing both opportunities to gain further insights into brain functionality, and a challenge on the analysis side; accuracy and efficiency. In the presented work, we showed that efficiency could be improved with some (arguably marginal) penalty on a range of popular machine learning accuracy outcomes that are applied for the analysis of EEG data. The introduced method assumes that much of EEG signal information can be captured by (A) signals in a subset of EEG channels, which we filtered by the application of mRMRe technique and (B) signal information from discrete time-series slices (3secs) which contain signal (eye-state) transitions. Experimental results obtained show that both assumptions hold for the classification of eye-state from 14 EEG channels based on the dataset provided by Roesler, see Section \ref{sec:Data}. 

When developing this work, results should be considered alongside deep neural network architectures which have shown low convergence times and may be applicable in real-time classification of eye-state \cite{Reddy:2016aa}. 

Slicing frequency and the number of features/channels to select have been done heuristically here, and therefore, based on these preliminary outcomes, we hope to validate the presented method and obtained outcomes on larger datasets, in future work. Also, we believe that more noticeable gains in learning efficiency should be possible for  datasets of significantly higher-dimensional features' spaces.

\section*{Code availability}
Authors will provide link to code and dataset in camera-ready manuscript. %%
\label{sect:bib}
\bibliographystyle{ACM-Reference-Format}
\bibliography{AsquithBib}


\end{document}